\documentclass{ws-procs9x6}
\graphicspath{{./}}
\begin{document}

\begin{center}
Accepted for publication in the Proceedings of the ICATPP Conference on
Cosmic Rays for Particle and Astroparticle Physics,\\ Villa  Olmo (Como, Italy), 7--8 October, 2010, \\to be published by World Scientific %(Singapore).
\end{center}
\vspace{-1.5cm}

\title{Electron and Positron solar modulation\\ and prediction for AMS‐02.}

\author{P. Bobik$^{1}$, M.J. Boschini$^{2,4}$, C. Consolandi$^{2} $, S. Della
Torre$^{2,5}$*,\\ M. Gervasi$^{2,3}$, D. Grandi$^{2}$, K.
Kudela$^{1}$, S. Pensotti$^{2,3}$, and P.G. Rancoita$^{2}$}

\address{$^{1}$ Institute of Experimental Physics, Kosice (Slovak Republic) \\
$^{2}$Istituto Nazionale di Fisica Nucleare, INFN Milano-Bicocca, Milano (Italy) \\
$^{3}$Department of Physics, University of Milano Bicocca, Milano (Italy) \\
$^{4}$CILEA, Segrate (MI) (Italy) \\
$^{5}$Department of Physics and Maths, University of Insubria, Como (Italy)\\
*E-mail: stefano.dellatorre@mib.infn.it}

\hyphenation{pa-ra-me-ter}
\hyphenation{mea-su-re-ments}
\hyphenation{con-se-cu-ti-ve}

\begin{abstract}
The solar modulation, a combination of diffusion, convection,
magnetic drift and energy loss inside the heliosphere is usually
seen as a depletion in the Galactic cosmic ray (CR) flux at low
energy (less than 10 GeV/nuc). Antiparticles such as antiprotons
or positrons undergo the same processes of respective particles
but with a different magnitude depending on the Solar magnetic
field polarity. For electrons and positrons, due to the small
mass, energy loss mechanisms as inverse compton, synchrotron,
bremsstrahlung and ionization have to be taken into account,
together with the typical adiabatic losses considered in the
heliosphere. We developed a Monte Carlo stochastic simulation with
the aim to compare the solar modulation of particles and
antiparticles in the same observation period. We are able to
estimate the different behaviours associated to the charge sign
dependent processes of the heliospheric modulation. We compared
the simulated positron fraction with measurements performed by
AMS-01 and PAMELA. We also present the prediction for the AMS-02
experiment.
\end{abstract}


\bodymatter

\section{Introduction}\label{intro}

Solar modulation of Cosmic Rays (CRs) is actually described by
model integrating the Parker's equation~\cite{parker1965}\!. The
CRs propagation is depending on the charge sign of CRs and to the
solar magnetic polarity~\cite{artmidrift,clem}\!. In this work we
introduce a description of our present modulation model, then we
summarize the different behaviours with particles of opposite
charge sign in same observation periods. Finally we compare
results with experimental data providing also a prediction for
AMS-02.

\section{Model Description}\label{model}

Present model~\cite{DavArticle2010}\! is based on the 2D (radius
and colatitude) approximation of Parker
equation~\cite{parker1965}\!: a Fokker-Planck equation including
diffusion, convection, adiabatic energy loss and magnetic drift. A
rigorous proof by Ito~\cite{gardiner1985}\! demonstrate the
equivalence with a set of ordinary stochastic differential
equations (SDEs). The main advantage is that SDEs could be easly
integrated with Montecarlo techniques. The model depends on some
parameters related to the solar activity and is fine-tuned by
comparing results with experimental data\cite{ICRC09}\!. We used
the Heliosferic Magnetic Field (HMF) introduced by Parker and
modified according to Jokipii and K\'ota~\cite{jokipii89}\!: the
magnitude of the HMF in the polar regions is increased without
modifying the topology of the field. The major effect is a reduced
CR penetration along the polar field lines in the inner part of
the heliosphere, due to a lower magnetic drift velocity.
\mbox{According} to Ulysses data~\cite{ulysses}\!, for periods of
low solar activity, we used a latitudinal dependence of the solar
wind speed: the value increases by a factor of two moving from the
ecliptic plane to the polar region. For the drift description we
use the model by Potgieter \&
Moraal~\cite{Potgieter85,DellaTorre2009} that describes periods of
solar maximum as well solar minimum. The model requires as input a
Local Interstellar Spectrum (LIS). We use the model proposed by
Zhang \& Cheng~\cite{zhang2001}\!, which has been corrected
comparing it with the experimental data~\cite{mauroPoster2010}\!.

\section{Results}

\subsection{Drift effect on modulation}

\begin{figure}[htp]
\begin{center}
 \includegraphics[width=0.48\textwidth]{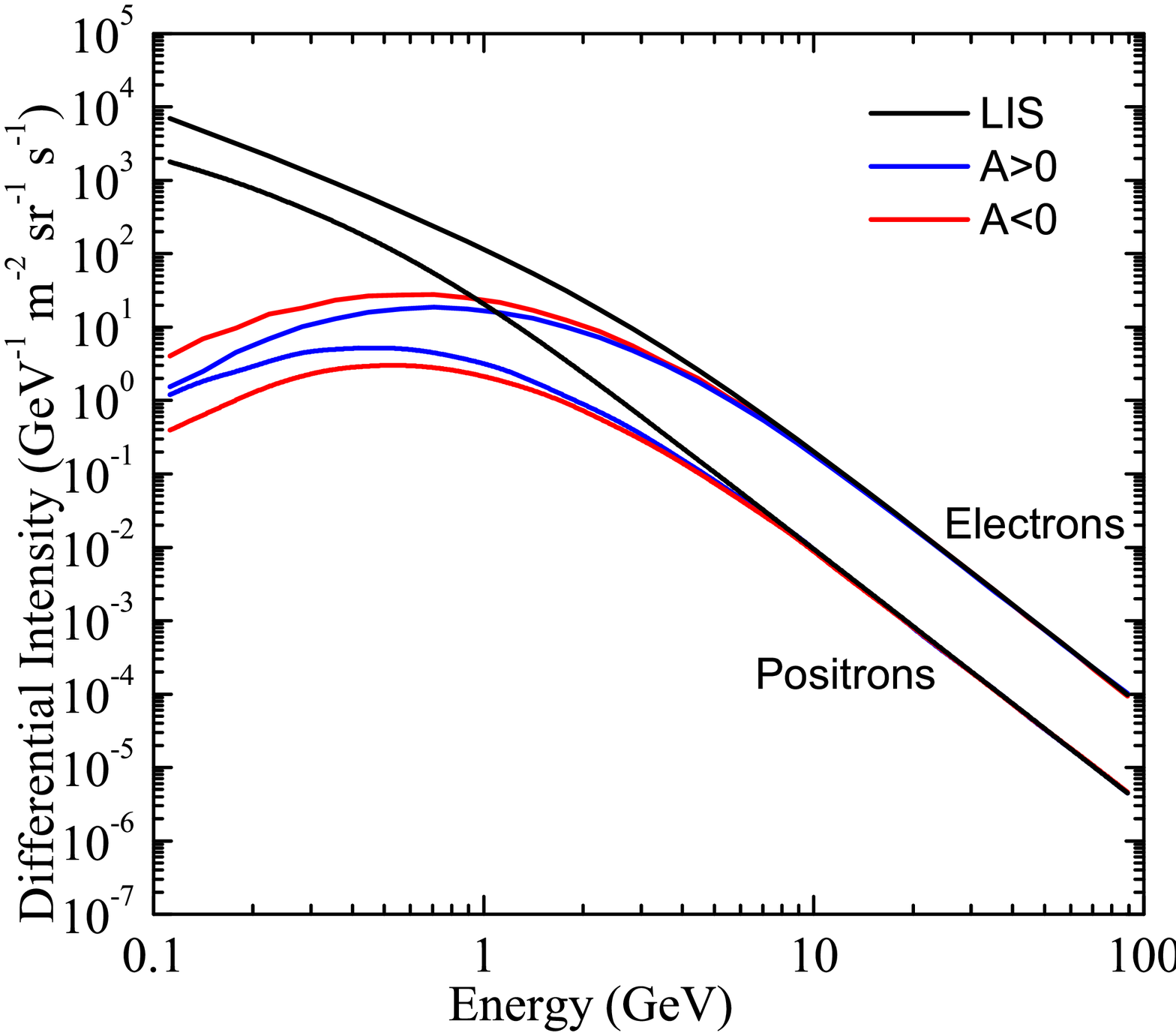}
 \includegraphics[width=0.48\textwidth]{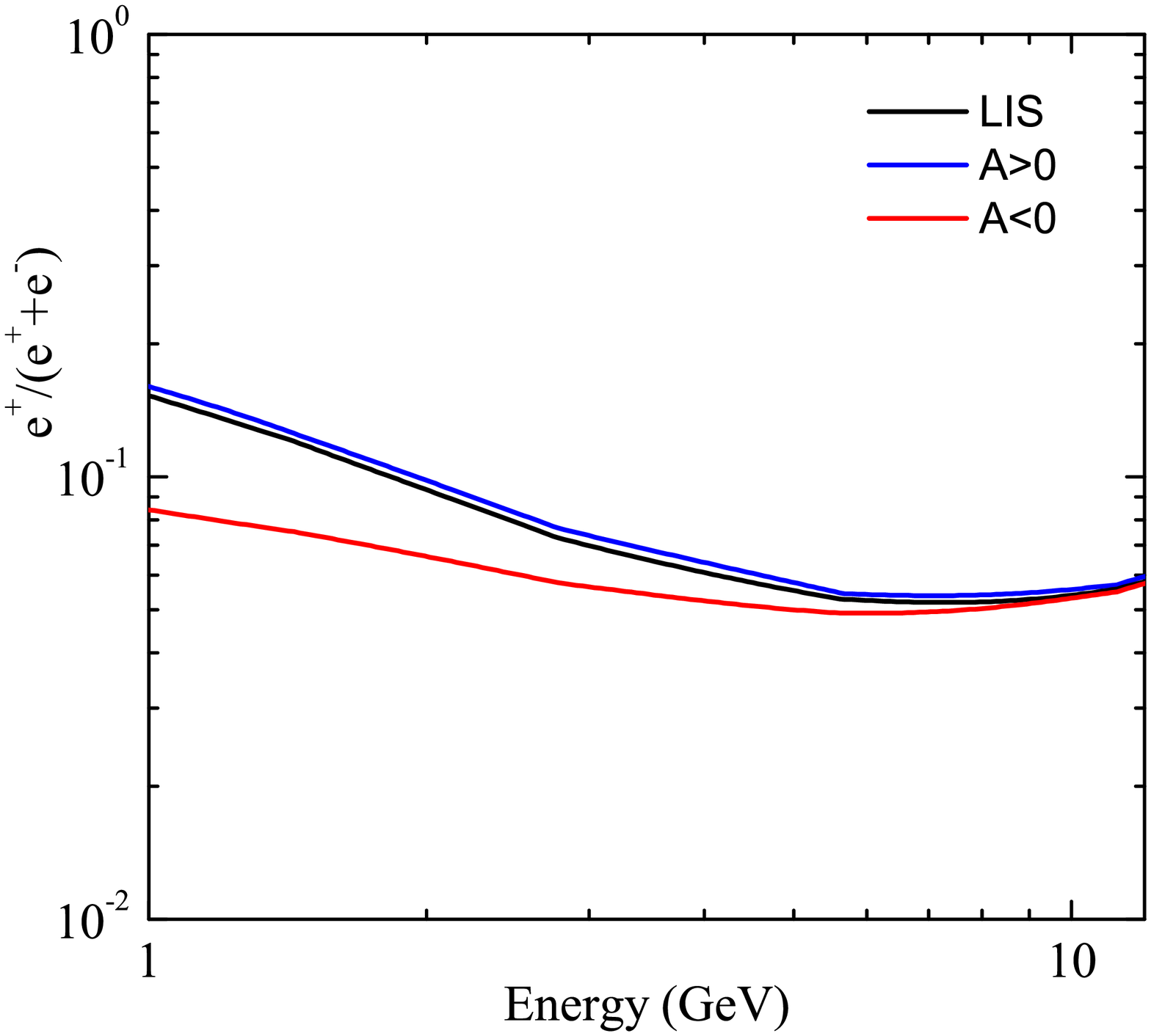}
 \caption{Electron and positron modulation (left) and positron ratio for a
 typical solar minimum period (we use conditions related to June 1998)
 with both magnetic field polarities.}\label{qAI}
\end{center}
\end{figure}

The charge dependence of propagation is due to the drift term of
the Parker equation.
%that is proportional to $\mathbf{v}_d=\nabla \times
%(\frac{pv}{3qB^2}\mathbf{B})$\cite{Hattingh95};
The fundamental parameter that fix the direction of particle
magnetic drift is $qA$, where $q$ is the particle charge and $A$
is the normalization factor of the magnetic field ($B$) that gives
its polarity. We made a simulation of a typical solar minimum in
both polarities for both electrons and positrons. We found, in
agreement with theory~\cite{alanko2007}\!, that the flux is lower
when $qA<0$ than for $qA>0$ conditions. Results are shown in fig.
\ref{qAI} (left panel). In the right panel of fig. \ref{qAI} we
show the modulated positron fraction above 1 GeV: for a period
with $A>0$ the modulation does not affect $e^- / (e^+ + e^-)$
ratio significantly; for a period with $A<0$ the effect of solar
modulation is relevant up to $\sim$ 10 GeV, reducing the modulated
fraction by a factor of two or more.

\subsection{Measurements by AMS-01 and PAMELA}
We selected two data set in similar solar conditions but opposite
\mbox{magnetic} field polarity. AMS-01\cite{amsPositron} operated
on board the Space Shuttle in June 1998, at the end of a solar
minimum occured during a period with $A>0$.
PAMELA\cite{PamelaPositron2009} is a space born experiment working
since July 2006. Published data have been taken between 2006 and
December 2008, therefore during the last long solar minimum with
$A<0$.

\begin{figure}[htp]
\begin{center}
 \includegraphics[width=0.48\textwidth]{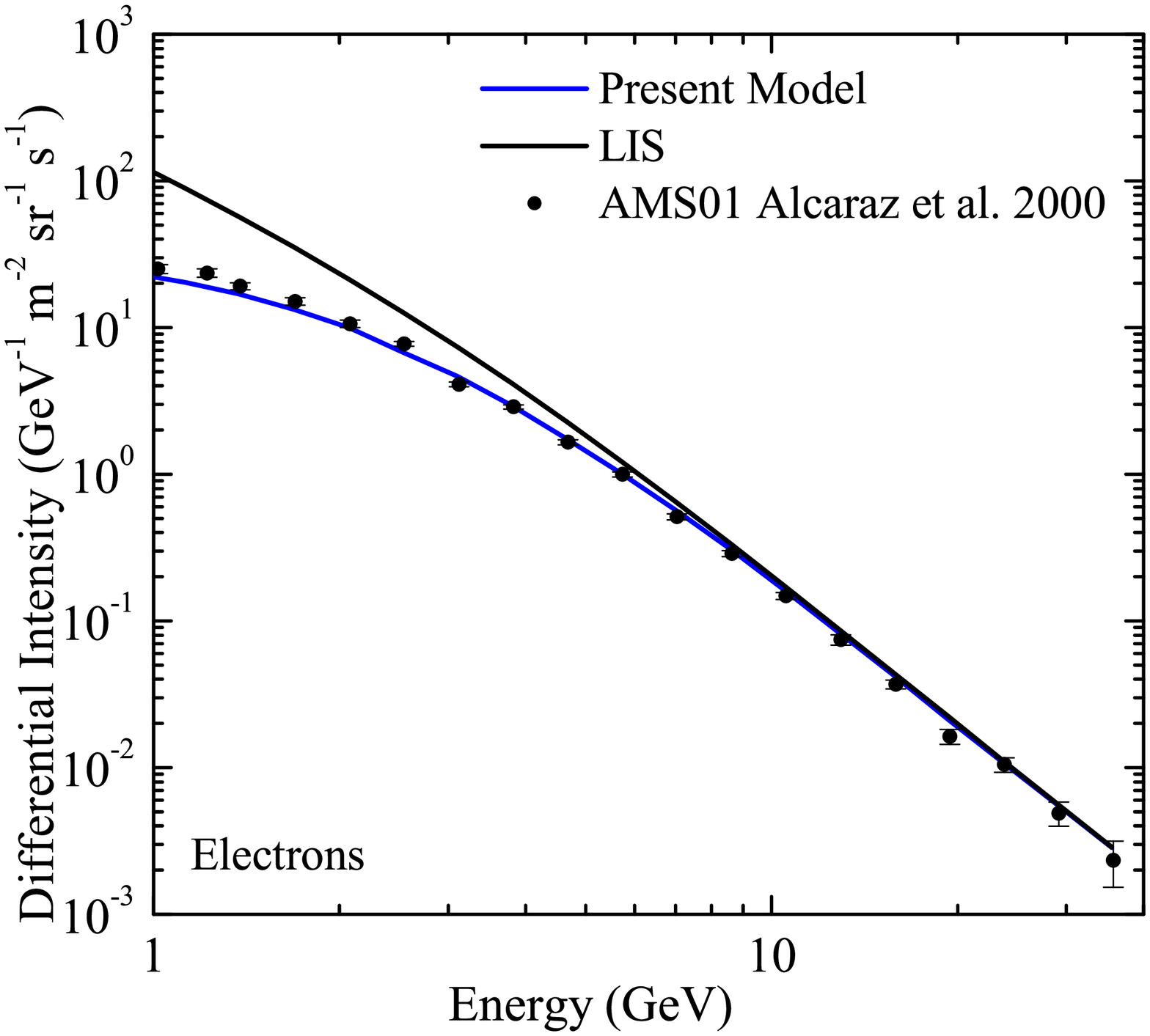}
 \includegraphics[width=0.48\textwidth]{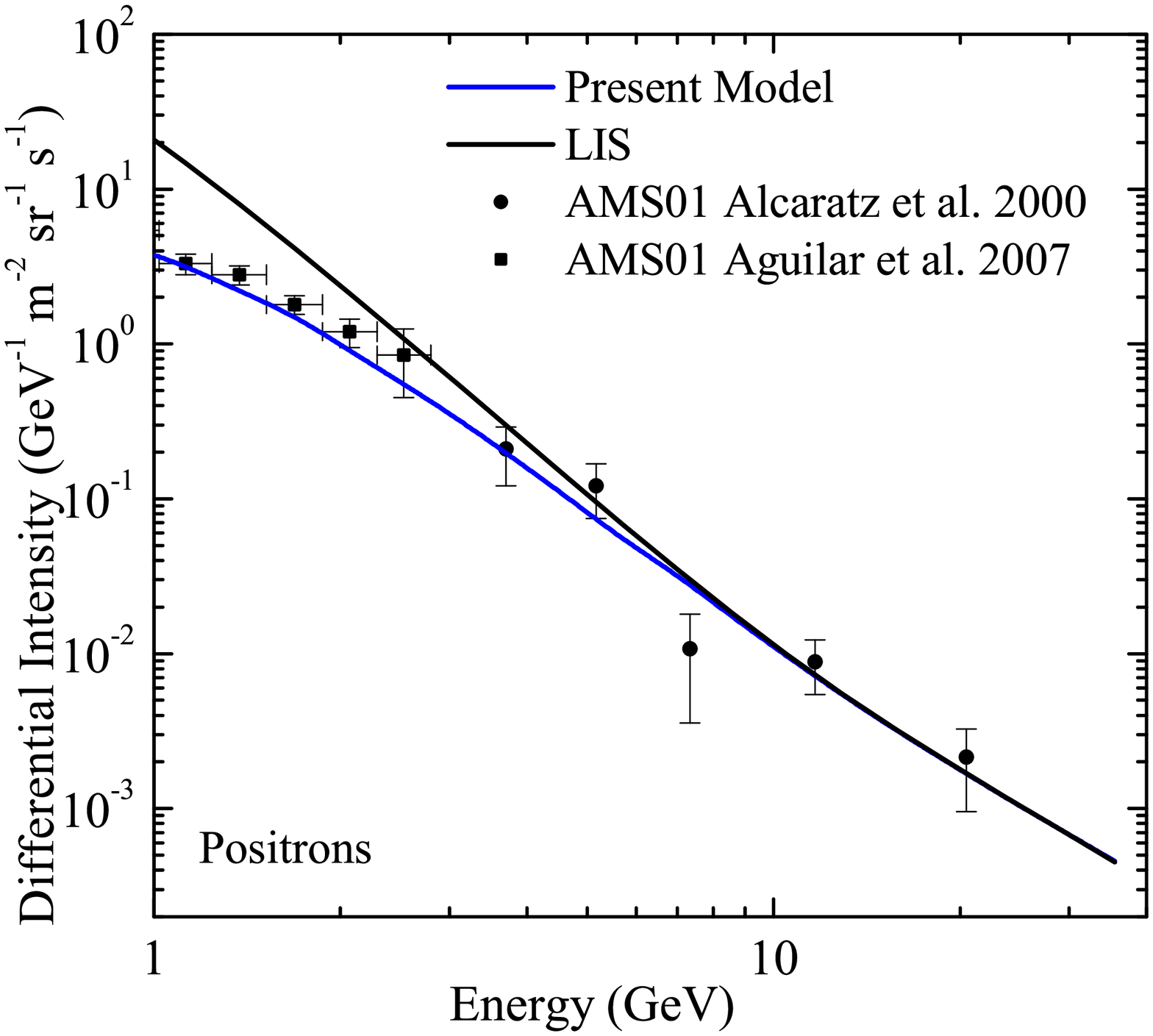}
 \caption{Simulated spectra for AMS-01 mission.}\label{AMSlepton}
\end{center}
\end{figure}

\begin{figure}[htp]
\begin{center}
 \includegraphics[width=0.48\textwidth]{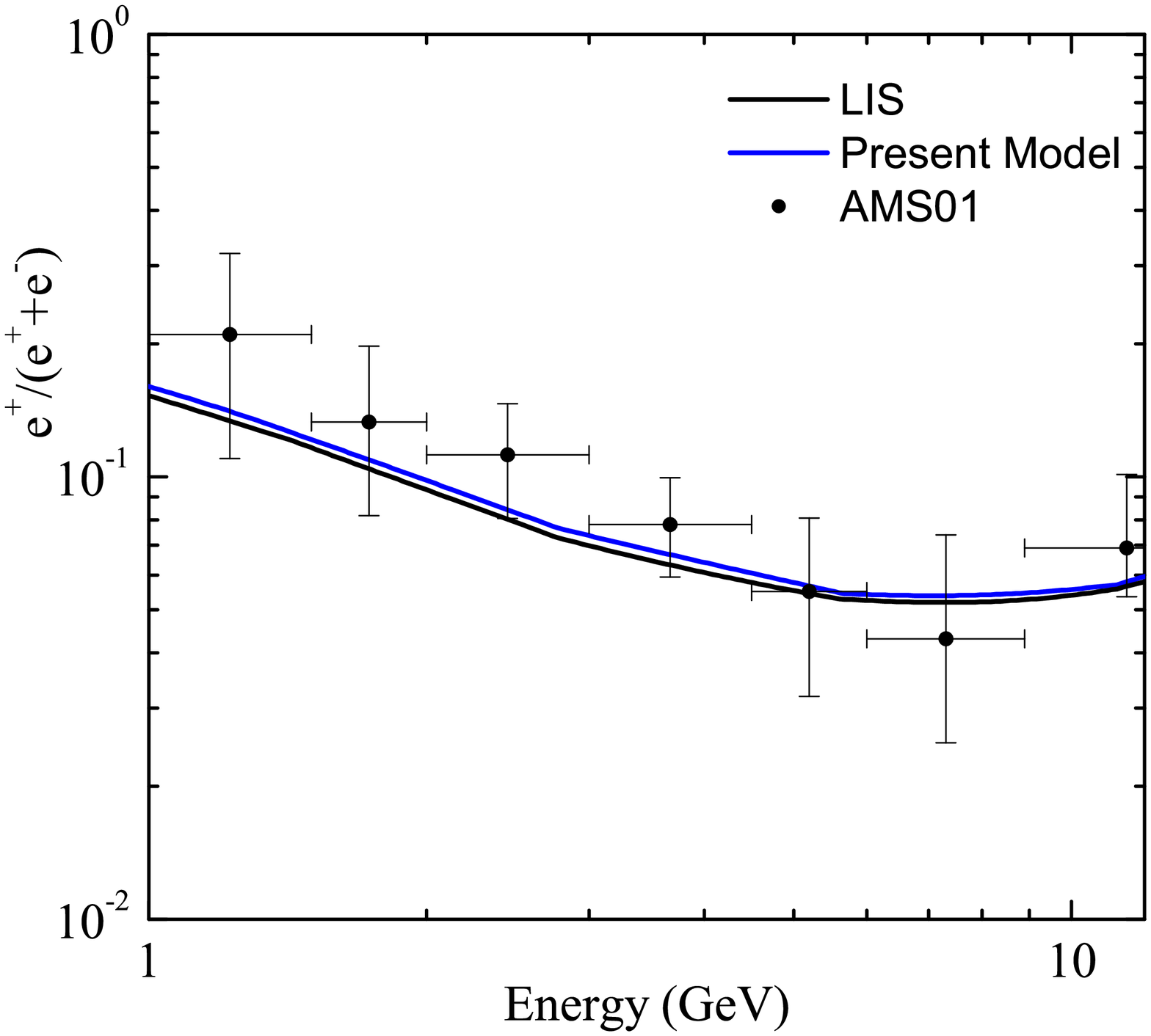}
 \includegraphics[width=0.48\textwidth]{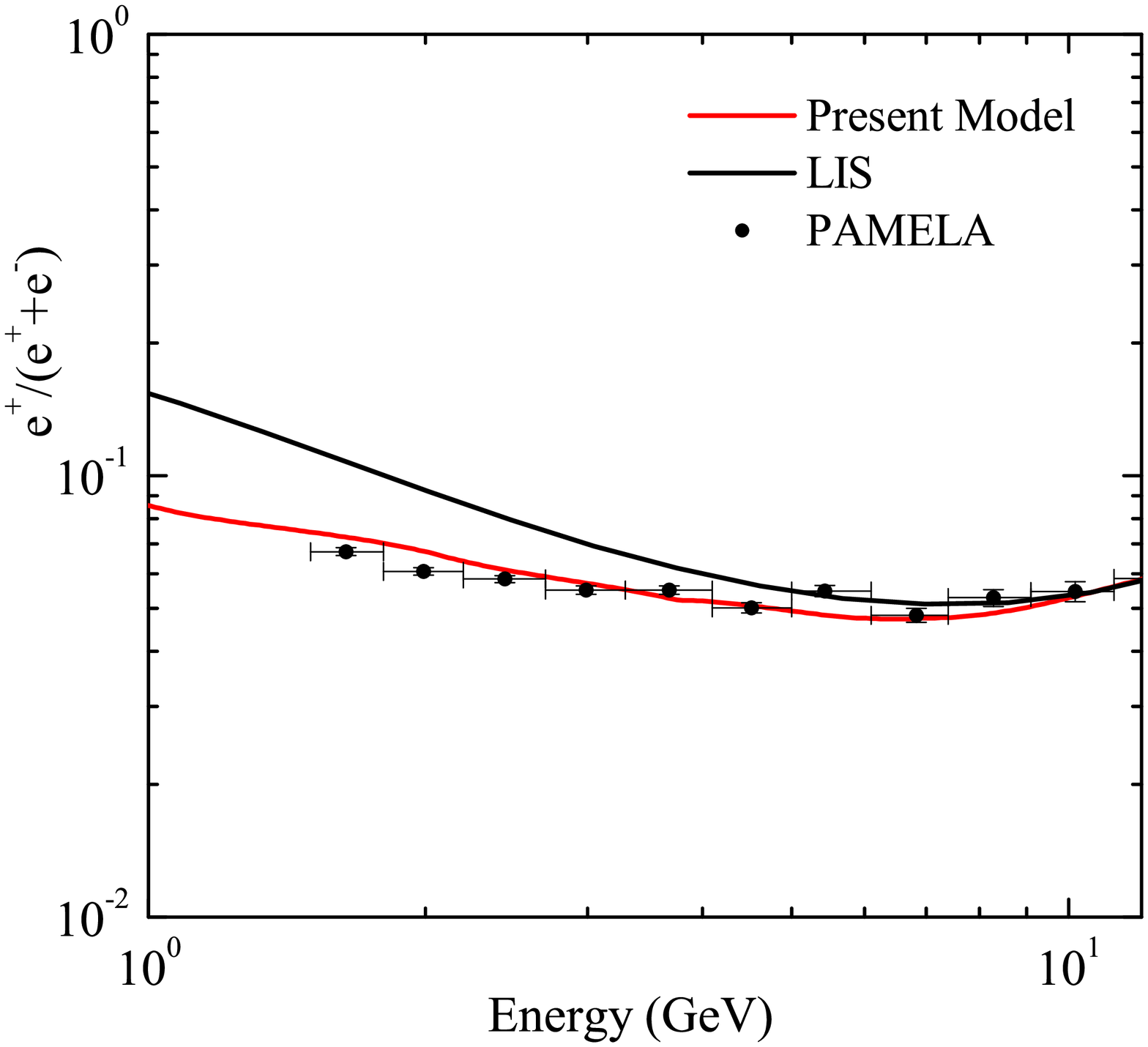}
 \caption{Simulations of positron ration for AMS-10 and PAMELA mission.}\label{RatioAMS}\label{RatioPamela}
\end{center}
\end{figure}

We reproduce, within the errors bars, the AMS-01 data for
electrons and positrons (see figure \ref{AMSlepton}). We find a
good agreement between simulation and experimental data also for
$e^+$ fraction. Despite of the large solar modulation of electron
and positron flux the modulated ratio is very close to the
interstellar ratio (figure \ref{RatioAMS}) as obtained in previous
results. We used the same model of propagation to reproduce data
published by PAMELA collaboration. We simulated several
heliospheric conditions covering the periods of the data taking.
The results have been averaged and shown in figure
\ref{RatioPamela}. We reproduced PAMELA data and find that
modulated ratio is lower than the interstellar one, as expected.

\subsection{Prediction for AMS-02}
\begin{figure}[htp]
\begin{center}
 \includegraphics[width=0.48\textwidth]{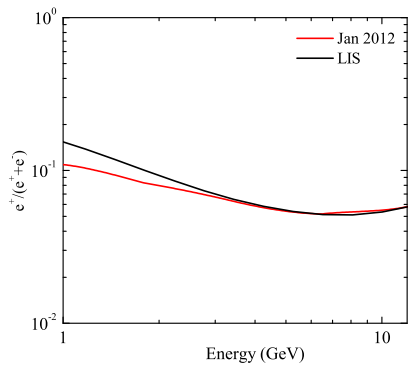}
 \caption{Prediction of positron ratio for January 2012.}\label{Ratioams02}
\end{center}
\end{figure}

The simulation code was used to predict CR fluxes for future
measurements. The periodic behavior of the heliosphere allows us
to predict, with a certain level of precision, the solar
modulation parameters needed for the simulation. The assumption is
that diffusion coefficient, tilt angle and solar wind speed show a
near-regular and almost periodic trend. The periodicity is two
consecutive 11-years solar cycles. In order to get these data we
considered the prediction of Smoothed Sunspot Numbers from IPS
(Ionospheric Prediction Service) of the Australian Bureau of
Meteorology. Using SIDAC data (Solar Influences Data Analysis
Center) we selected periods with a similar solar activity
conditions and same solar field polarity of the simulation time:
therefore approximately 22 years before. Under the previous
assumpion, we used the values measured in that periods as an
estimation of the conditions of the heliosphere in the near
future. We concentrate the simulations on the AMS-02 mission that
will be installed on the ISS in February 2011, and, in particular,
on one period, during its data taking, approaching the solar
maximum: January 2012. results are shown in figure
\ref{Ratioams02}. This period is still with $A<0$ and the
modulated positron fraction is still below the interstellar one.

\section{Conclusions}\label{conclusion}

We built a 2D stochastic Monte Carlo code for particles
propagation across the heliosphere. Present model takes into
account drift effects and shows quantitatively a good agreement
with measured values, both for positive and negative periods and
for different particle's charge. We found that for periods with
$A>0$ the modulated positron ratio is similar to the interstellar
one. During periods with opposite field polarity we found instead
a significative reduction of the modulated positron ratio respect
to the interstellar ratio. We compared simulations with positron
ratio measured by AMS-01 and PAMELA. We found that the discrepancy
of PAMELA positron ratio respect to the previous experiments can
be well explained with a polarity-dependent effect of modulation,
due to magnetic drift of particle. The prediction for AMS-02 shows
that, in the first period of data taking, it will probably observe
a modulation effect similar to the one measured by PAMELA. It will
be interesting to compare the measurements of AMS-02 in the
following years to confirm the behavior shown in fig. 1.

\bibliographystyle{ws-procs9x6}
%\bibliography{ws-pro-sample}
\bibliography{DellaTorre_ICATPP}

\end{document}